\documentclass[review]{elsarticle}
\usepackage{algorithm}
\usepackage{multicol}
\usepackage{algorithmic, amsmath}
\usepackage{amsfonts}
\usepackage{epsfig,graphicx}
\usepackage{enumerate}
\usepackage{calligra}
\usepackage{longtable}
\usepackage{mathptmx}       
\usepackage{helvet}         
\usepackage{courier}        
\usepackage{type1cm}        
\usepackage{makeidx}         
\usepackage{graphicx}        
\usepackage{multicol}        
\usepackage[bottom]{footmisc}

\usepackage{amsthm}
\usepackage{adjustbox}
\usepackage[figuresright]{rotating}
\theoremstyle{plain}
\newtheorem{thm}{Theorem}[section]
\newtheorem{lem}[thm]{Lemma}

\theoremstyle{definition}

\begin{document}

\begin{frontmatter}
\title{ Fair End to End Window Based Congestion Control in Time Varying Data Communication Networks}
\author{Suchi Kumari}
\address{Department of Computer Science and Engineering,\\
National Institute of Technology,\\
Delhi, India\\ email: {suchisingh@nitdelhi.ac.in}}
\author{Anurag Singh}
\address{Department of Computer Science and Engineering,\\
National Institute of Technology,\\
Delhi, India\\email: {Corresponding author: anuragsg@nitdelhi.ac.in}}

\begin{abstract}
Communication networks are time-varying and hence, fair sharing of network resources among the users in such dynamic environment is a challenging task. In this context, a time-varying network model is designed and shortest user's route is found. In the designed network model, an end to end window-based congestion control scheme is developed with the help of internal nodes or router and the end user can get implicit feedback (RTT and throughput). This scheme is considered as fair if the allocation of resources among users minimizes overall congestion or backlog in the networks. Window update approach is based on multi-class fluid model and is updated dynamically by considering delays (communication, propagation and queuing) and the backlog of packets in the user's routes. Convergence and stability of the window size are obtained using a Lyapunov function. A comparative study with other window-based methods is also provided.
\end{abstract}


\end{frontmatter}

\section{Introduction}
Real life systems such as communication, social, biological can be described with the help of complex networks \cite{newman2011complex,COMPLX}. It is a challenging task to design a time-varying communication network (TVCN) with the ability to respond to the randomly changing traffic. Hence, the study of the traffic dynamics and fair sharing of resources on communication networks has received a great wave of interest for the researchers in past few years. The allocation of resources among the users in an unbiased manner is one of the challenging tasks in today's scenario. For assigning resources in an unbiased or fair manner, some researchers use various rate vector allocation schemes \cite{kelly2001,la2002} to gain maximum utility while others select fair end to end window-based congestion control scheme \cite{mo2000fair,la2002}. In this paper, we are interested to design a TVCN model considering network growth, redistribution of traffic from heavily loaded nodes and removal of some fraction of links to reduce maintenance cost. A window-based congestion control scheme may be applied on the proposed TVCN model and user's current window size may be updated by considering delays(communication, propagation and queuing delays). 

Communication networks are evolving and the concept of evolving networks with preferential linking during the addition of new nodes is introduced by Barabasi-Albert (\cite{bara}). The distribution of degrees of nodes in these networks follow the power law and is termed as scale free nature of the networks. Many time-varying graph (TVG) models are proposed \cite{TVG,TVGdynamic}. A series of static graphs (i.e., the snapshots) is used to represent the network at a given time instant \cite{snapshot}. Wehmuth \textit{et al.} \cite{TVG} proposed a new unifying model for representing finite discrete TVGs. A framework is designed to obtain degree distribution of evolving network with the consideration of deletion of nodes and a continuum formulation is also provided by \cite{2007continuum}. A preferential attachment model for network growth is proposed where a new node has partial information about the network \cite{2015partial}. A new node has access to a fraction of nodes and a new connection is formed with the known set of nodes with a probability proportional to the degree of the node. A framework to represent mobile networks dynamically in a Spatio-temporal fashion is designed and algebraic structural representation is also provided by \cite{kumari2016}. Chen \textit{et al.} \cite{chencontrollability} investigated controllability of a formation control systems in a directed graph where a node represents agent and link describes information flow between nodes in the system. They studied strong and weak connected components in a time-varying graph and provided some estimations on a number of nodes of strongly connected components.

Communication networks are extremely changeable and modeling its rate control behavior helps us in getting optimal utilization of the system. Kelly \cite{kelly2001} provided a mathematical model of the Internet where rate allocation problem is considered as an assignment of optimal rate to each user for maximizing individual user's utility as well as the system utility. La \textit{et al.} \cite{la2002} extended Kelly's work by introducing a suitable pricing scheme. Resources may be of multiple types and users request different portion of the resources so trade-off arises between fair resource allocation and system efficiency \cite{multiresource}. Anirudh \textit{et al.} \cite{experimental} have used an automated protocol design tool which approximates the best possible congestion control scheme without the prior information about the structure of the network.

Some Congestion control algorithms in the elementary transmission control protocol (TCP) based models such as TCP Reno \cite{2000Reno}, Tahoe \cite{mo1999analysis} and TCP Vegas \cite{brakmo1995tcp} have performed remarkably well. As current Internet scaled up by more than six times in size, speed and load in past few years. Hence, due to increased bandwidth delay, the model will eventually become a performance bottleneck. A new congestion control scheme, called FAST is developed by proposing an equation based algorithm to avoid packet level oscillation, using queuing delay as a measure of congestion and stable and weighed proportionally fair flow is obtained \cite{jin2005fast}. Proportional fairness is achieved by their $ (p,1) $-proportionally fair algorithm and max-min fairness is achieved as a limiting case of $ (p,\alpha) $ proportionally fairness. Mo \textit{et al.} \cite{mo2000fair} and La \textit{et al.} \cite{la} proposed a window based algorithm, in which window sizes are adjusted more intelligently such that transmission rate and the backlog of packets should be controlled. TCP/AQM \cite{low2003duality} based optimization technique performs well under commonly defined constraints but it is hard to simulate in the current complex networks. Low \textit{et al.} \cite{low2018} proposed a framework to cop up with buffer-bloat in multiple bottleneck links. The stability of the Controlled Delay (CoDel) is also analyzed and system stability and performance are improved by designing a self-tuning CoDel. In delay tolerant network (DTN), there is no guarantee of an end to end connection between source and destination hence, it is a challenging task to perform some functions like routing and congestion control. Aloizio \textit{et al.} \cite{silva2015survey} performed classification on existing DTN and provided a state of the art in DTN congestion control mechanisms. Giovanna\textit{ et al.} \cite{2016optimal} considered multi-path routing with joint congestion control and user throughput is maximized, and, overall network cost is minimized. 

Some evolving network models only consider the addition of nodes \cite{bara}, although some works consider deletion of nodes during/beyond the evolution of the networks \cite{chen2004,deletion,2007continuum}. There are only a few networks such as citation, movie actor collaboration and science collaboration networks which can be described without removal of links. But other real-world networks, e.g., like the Internet, communication networks, WWW, transportation networks have the varying composition of links due to appearance or disappearance of connections. A link may disappear due to break down and sometimes removed forcibly to balance the traffic loads in the communication networks. Few heavily loaded nodes get the opportunity to rewire (remove a connection from a node and attach with another node) their links. In this context, the proposed work considers designing of a TVCN model, minimization of network congestion and a fair sharing of resources among the users according to their willingness to pay. A TVCN is designed where nodes, as well as links, are getting added into the networks while restructuring is performed in the existing networks through rewiring and removal of the links. Addition and rewiring of links are based on preferential attachment while removal of links is based on anti-preferential attachment \cite{SK2017}. Probability $ \Pi_i $ that a node  $ i $ will be selected through preferential attachment is proportional to its in-degree and is given by, $ \frac{k_i}{\sum_{j \in N} k_j} $ where, $ k_i $ is in-degree and the probability $ \Pi'_i $ of selecting node $ i $ with anti-preferential attachment is given by, $ \frac{1}{|N|-1} \left(1-\frac{k_i}{\sum_{j \in N} k_j}\right) $. Rewiring of links is done to redistribute the traffic loads, while few links increase the maintenance cost so these links may be chosen for removal from the network. After designing the network, users select their routes for data communication with their destination nodes. Optimality of the routes is checked by using the proposed window-based congestion control scheme at different time intervals.

In the window based algorithm, a fair end to end congestion control scheme is developed with the help of internal node or router. Each user calculates congestion at the resources appearing in its route without using any explicit signal from the network router. It is hard to get prior information of network's feedback as networks are evolving over time. In this paper, congestion control formulation is considered as fair allocation which is a natural extension of the models proposed by Mo \textit{et al.} \cite{mo2000fair} and La \textit{et al.} \cite{la} and a comparative study of these models are also provided. Mo \textit{et al.} \cite{mo2000fair} updated window size by considering propagation delay, queuing delay and using the difference between the actual and the target backlog. La \textit{et al.} \cite{la} updated window size such that it converges to the unique stable point, where the resulting rates solve the system utility, $ SYSTEM (U, A, C) $. As networks keep changing with time and evolving in nature so it is required to consider dynamic parameters in the window updating rules. The propagation delay is a function of the distance between two nodes and remains constant throughout the life of the networks. Hence, in place of propagation delays, we have considered transmission delay of the resources. The transmission delay is the amount of time required for the node to push out the packet and it is dependent on the transmission rate of the connecting link and the length of the packet. Once user's connection is established, packets are transferred to the receiver and number of packets is decreased at any resource with some rates. Hence, the transmission delay also decreases accordingly.  

In this paper, Section 2 describes fairness and gives background information about the classical models. Section 3 states the proposed work in two parts: (i)the time varying communication network model and (ii) fair end to end congestion control scheme. Section 4 presents simulation results, and in Section 5, conclusions and future research plan are discussed.

\section{Fairness and Classical Models}
Network structure keeps on changing with time hence, nodes as well as links will change accordingly. Let the network consist of $ N $ nodes, $ E $ links (resources) and $ T $ is life time of the network. A set of $ R $ users are willing to access and send data to the desired location through the networks using the resources. A link $ e_{mn} $ establishes a connection between node $ m $ and node $ n $ and can send maximum $ C_{e_{mn}} $ units of data through it, where capacity of a link $ C_{e_{mn}} $ is dependent on the degrees $ k_m $ and $ k_n $ of nodes $ m $ and $ n $. Each user $ r $ is assigned a route $ r \in R $ in the duration of time $t_{(i-1)} \in T $ to $ t_i \in T $. At the end of $ t_i^{th} $ time, a zero-one matrix $ A $ of the size $( R \times E )_{t_i} $ is defined where, $ A_{r, e_{ab}, t_i} = 1 $, if link $ e_{ab} $ is a part of route  $ r_i $ in the duration of time $ t_{(i-1)} $ to $ t_i $, otherwise zero. 
 
\subsection{Fairness}
Fairness is defined as how the bottleneck resources are shared among users in communication networks. There are various types of fairness measures.
\begin{itemize}
\item Max-min fairness: A feasible flow vector is defined as max-min fair if any rate vector $ x_i $ cannot be increased without decreasing some $ x_j $ where, $ x_j \leq x_i $. The max-min fairness is defined in terms of transmission rates of source nodes sending data to destination nodes \cite{max}. We need transmission schedules of the packets along with the global state and timing information of all the nodes in the network hence, its use is avoided in dynamic communication networks.
\item Proportional fairness: A rate vector $ x^* $ is called proportionally fair if it is feasible. For any other feasible vector $ x $, the aggregate of the proportional change is negative.
\begin{equation*}
\sum_i \frac{x_i - x_i^*}{x_i^*} \leq 0
\end{equation*}
\item $ (p, \alpha)- $ Proportional fairness: There is a trade-off between fairness and resource utility maximization. Max-min fairness approach gives more priority on fairness. But, in real life scenario, we want to maximize overall network utility. Therefore, we need to generalize max-min fairness and proportional fairness. $ (p, \alpha)- $ defines both the fairness approach; proportional and max-min. If $ \alpha = 1 $, then it converges to proportional fairness otherwise it will behave as max-min fairness.
\end{itemize}

As we want to get optimal window size according to the willingness of pay $ p_r $ for each user, $ r $ so $ (p,1) $ proportional fairness approach may be used. The vector $ p $ in $ (p,1) $ is proportionally fair and dependent on utility function $U_r(x_r (t_i))$ 

\subsection{Fair End to End Window Based Model}
Window based model proposed by Mo \textit{et al.} \cite{mo2000fair} and La \textit{et al.} \cite{la}, is based on the fluid model of the network and is dependent on window size, rate and queue size. The model can be represented by,
\begin{eqnarray}
&& A^Tx-c \leq 0, \label{e3} \\ 
&& Q(A^T x - c) = 0, \label{e4}\\
&& X(Ad_Q + d_{prop}) = w, \label{e5}\\
&& x \geq 0, q \geq 0. \nonumber
\end{eqnarray}
Where, $ x = (1, ..., R)^T $, $ c = (1, ..., E)^T $, $ d_Q = (1, ..., E)^T $, $ d_{prop} = (1, ..., R)^T $ and $ X= diag(x) $. $ d_{prop} $ is propagation delay, $ d_Q $ is queuing delay and $ (Ad_Q + d_{prop}) $ is total delay. Eq. \eqref{e3} shows capacity constraint. The constraint in Eq. \eqref{e4} states that there may be backlogged data at any resource if total data rate is equal to its capacity. Window size of a connection is sum of the number of packets in transmission and the packets buffered in the queue (Eq. \eqref{e5}). Rate vector, $ x $ and queue size, $ q $ should be positive. 

Mo \textit{et al.} \cite{mo2000fair} considered $ (p,1) $ proportionally fair algorithm where $ p $ is the vector of target queue size of the connections. Let $ ((Ad_Q)^i + d_{prop}^i), d_{prop}^i, w_i(t) $ and $ x_i(t) $ denote the total delay, propagation delay, current window size and data rate of user $ i $, respectively. They update user's window size based on the following system of differential equations,
\begin{eqnarray}
&& \frac{d w(t)}{dt} = -\alpha \frac{d_{prop}^i}{((Ad_Q)^i + d_{prop}^i)} \frac{s_i(t)}{w_i(t)}, \label{e6}\\ 
&& s_i = w_i - x_i d_{prop}^i -p_i. \nonumber
\end{eqnarray}
Where, $ s_i $ is the difference between actual and targeted backlog of user $ i $.

La \textit{et al.} introduced dynamicity in the above formulation, where $ p_i $ is dependent on $ SYSTEM(U(t_i),A(t_i)) $ and $ d_Q $ is updating dynamically with time \cite{la},
\begin{eqnarray}
&& \frac{d w(t)}{dt} = -\alpha \frac{d_{prop}^i + U'_i(x_i(t))+ x_i(t)U''_i(x_i(t))}{((A d_Q)^i(t) + d_{prop}^i)} \frac{s_i}{w_i}. \label{e7}
\end{eqnarray} 
As $ U_i(x_i(t)) $ is an increasing concave function. Hence, the term $ U'_i(x_i(t))+ x_i(t)U''_i(x_i(t)) $ in Eq. \eqref{e7} will always be positive and depend on the rate and user's utility and is added into the propagation delay. It shows that user's window size update equation depends on user's utility function. 
\vspace*{-0.5cm} 
\section{Proposed Work}
We want to find the node which is endorsed by the maximum number of nodes. Hence, the probability of linking of a newly appeared node with nodes with a maximum degree will be greater. The network is shared among users and each user wants to maximize its system utility and minimize congestion in the network by choosing an optimal window size. Therefore, the formulation of a congestion control scheme for a dynamic network is also provided.
\subsection{Time Varying Communication Network Model}
A time-varying communication network model is proposed where, at each time instant, $ t_i \in T $, a new node $ i $ is added to the network (expansion) and a number $ M (\leq n_0) $ is selected for network expansion, rewiring, and removal, where $ n_0 $ is initial number of nodes present as the seed network. Links are divided into three categories: newly added, rewired and removed links. Distribution of the links are done using the given set of rules \cite{SK}.
\\
\textbf{Notations:}
\begin{enumerate}[(a.)]
\item $ \beta =$ Fraction of the evolving (appear from the new node and appear/disappear in the existing network) links at any time instant. It informs about the establishment of new connections from the new nodes at time $ t = n $ ,  $ 0 < \beta < 1 $.
\item $ \gamma= $ Fraction of the links, rewired in the existing network, $0.5 < \gamma \leq  1 $.
\end{enumerate}
Using the above notations, following set of rules are formed,
\begin{enumerate}[(i)]
\item Total number of new out-flowing links from the new appeared node, $ i \mbox{  }(=t) $ with the nodes in the existing network at $ (t-1) $ based on the preferential attachment is given by,
\begin{equation*}
 f_{add} (t) = \beta M 
\end{equation*}
\item Few links are rewired in the existing network, $ \gamma $ fraction of the available $ M $ links is chosen for rewiring,
\begin{equation*}
 f_{rewire}(t) = \gamma (M - f_{add}(t)) = \gamma (1-\beta) M 
\end{equation*}
\item Final fraction of the remaining segment of $ M $ are used for deleting the most infrequently used links.
\begin{equation*}
f_{delete}(t) = M- f_{add}(t) - f_{rewire}(t) = (1-\gamma)(1-\beta) M
\end{equation*} 
\end{enumerate}

\subsection{Proposed Fair End to End Window Based Model}
A TVCN is designed by using the proposed model. An end to end connection is established between user's source and destination. There exist multiple paths from user's source to destination nodes but among all those connections, the shortest route(s) is (are) chosen for data communication. Initially, user $ i $ wants to send $ w_i $ number of packets but an actual number of packets in user's route are the sum of packets in transit (on the route) and the total number of packets waiting in the queue. At each time instant, few packets are forwarded towards their destination node hence, each user updates its current window size. Users' preferences are implicitly reflected in the window size updating rule. An end to end window-based congestion control formulation is required to get a fair rate vector. Max-min fairness is not suitable for time-varying networks and it is more focused on fairness rather than the system utility maximization. In real-world networks, we are interested to maximize the system utility. Hence, $(p,\alpha)$ proportionally fair allocation for rate vector of each user is considered. Here, the value of $ \alpha  $ may be any positive number. 

\begin{figure}[!htb]
\begin{center}
\includegraphics[width=0.8\linewidth, height=2 in]{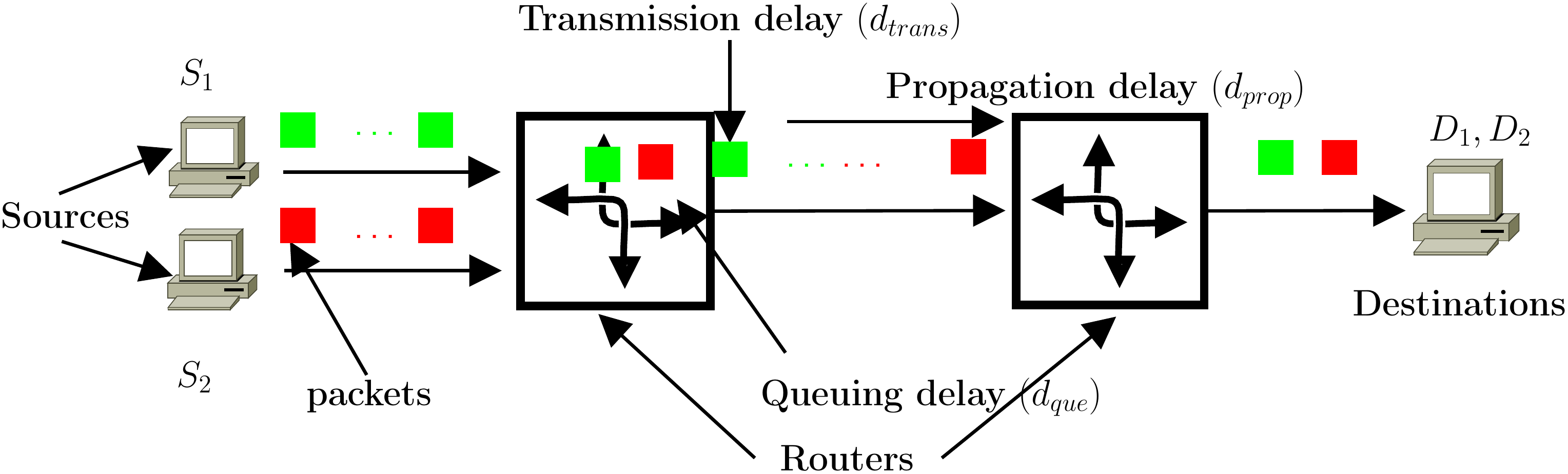} 
\caption{ Network Delays }
\label{f1}
\end{center} 
\end{figure}

Delays play an important role in finding congestion in the network as congestion is directly proportional to the delays. There are various types of delays in communication networks: transmission delay, propagation delay, queuing delay etc. A snapshot of all the delays are shown in Figure \ref{f1}. The transmission delay is defined by the amount of time required for the node to push out the packet. If the length of each packet is $ \mathtt{L} $ bits, and the transmission rate of the link from first node to second node is $ r_{tran} $ bits/sec, then transmission delay $ (d_{trans}) $ is denoted by, $ d_{trans} = \frac{\mathtt{L}}{r_{tran}} $ and it depends upon transmission rate $ (r_{tran}) $ of the link and length, $ L $. The propagation delay $ (d_{prop}) $ is the time to propagate a bit from one node to other. If the distance between two nodes is $ \delta $ and the propagation rate is $ r_{prop} $, then the propagation delay is $ d_{prop} = \frac{\delta}{r_{prop}} $. It is a function of the distance between the two nodes, but does not share the dependency on $ L $ and $ r_{tran} $ of the link. Queuing delay $ (d_Q) $ depends on number of packets waiting in a queue. Total delay $ D^i $ for the user $ i $ is the sum of propagation delay  $ d_{prop}^i $, transmission delay $ d_{trans}^i $ and queuing delay $ d_Q^i $ for that user.

It is hard to know for end users about the fair share of the network resources among the users. The fair share not only depends on the users but also the network structure. Each user does not aware of the behavior of other users. Hence, it is important to introduce a formulation where the user can achieve fairness without considering the behavior of other users. Let, $ w_i(t), x_i(t), d_{trans}^i(t) $ and $ p_i(t) $ represent the window size, data sending rate, transmission delay and willingness to pay of user $ i $, respectively at time $ t $. The actual backlog at user $ i $ may be defined as,
\begin{equation*}
s_i(t) = w_i(t)- x_i(t) d_{trans}^i(t) - p_i(t), \mbox{ for } i \in R.
\end{equation*}
Where, second term on the right-hand side of the equation provides the total number of packets transmitted through the route of user $ i $ in one round. $ w_i(t) $ is the total number of packets in the route of user $ i $. $ p_i(t) $ is willingness to pay for user $ i $ at time $ t $. A time varying cost is chosen by the user $ i $ for the queue size of resources which appears in the route of that user at time $ t $. User $ i $ is able to change its willingness to pay $ p_i(t) $ according to its rate $ x_i(t) $ as,\begin{equation}
p_i(t) = x_i(t) U'_i(x_i(t)), \mbox{ for } i \in R.
\end{equation} 
Where, $ U'_i(x_i(t)) $ is the first derivative of the utility function and is used to check whether the utility function is increasing or decreasing. Suppose a resource is heavily loaded with packets then the charge per unit flow through that resource will be high and user has to pay more. If user pays more then it may get high data sending rate, $ x_i(t) $. Hence, rate allocation is proportional to $ p_i(t) $ and is considered as $ (p,1) $- proportionally fair. The expression $ x_i(t) d_{trans}^i(t) - p_i(t) $ states the sum of the packets of user $ i $ which can send through the route and the total number of packets of user $ i $ which may store in the queue at the destination node. Therefore, the expression,  $ w_i(t) - x_i(t) d_{trans}^i(t) - p_i(t) $ will provide the total number of backlogged packets in the route of user $ i $. The choice for selecting the window size, $ w_i(t) $ of user $ i $ which makes $ s_i(t) \approx 0 $ is considered as $ (p_i,1) $-proportionally fair. User $ i $ needs to send $ w_i(t) $ packets while total delay is $ D^i(t)) $ hence, total time taken is $ w_i(t) D^i(t) $. But, total time to transmit backlogged packets, $ s_i(t) $ in one round is $ s_i(t) d_{trans}^i(t) $. Therefore, the fraction of packets that may be sent in one round is $ \frac{d_{trans}^i(t) s_i(t)}{D^i(t) w_i(t)} $. The rate equation that is used by the user to update its window size is,
\begin{equation}
\frac{dw_i(t)}{dt} = - \kappa f_i(d,t) f_i(w,t). \label{e8}
\end{equation}
Where,
\begin{equation*}
f_i(d,t) = \frac{d_{trans}^i(t)}{D^i(t)},
\end{equation*}
here, $ \kappa $ is a scaling parameter, $ \kappa > 0 $. Queuing delay, $ d_Q^i(t) = \sum_{i : j \in R_i} \left(\frac{\sum_{z : j \in R_z} x_z(t)}{C_j}\right) $ depends on number of packets in the queue and the route of user $ i $. If the incoming rate of the packet is $ \lambda_j(t) $ bits  at node $ j $, and the transmission rate of the user $ i $ is $ min_{1 \leq k \leq |R_i|} C_j $ bits/sec, then transmission delay $ (d_{trans}) $ is denoted by $ d_{trans}^i(t) = \frac{\sum_{j \in R_i} \lambda_j(t)}{min_{1 \leq < k \leq |R_i|} C_j}  $ and function $ f_i(w,t) $ is defined as,
\begin{equation*}
f_i(w,t) = \frac{w_i(t) - x_i(t) d_{trans}^i(t) - x_i(t) U'_i(x_i(t))}{w_i(t)}.
\end{equation*}
If resource (link) $ j $ appears in the route of large number of users then the number of packets at that resource will increase rapidly which makes the node congested. Hence, transmission delay and queuing delay will increase accordingly.

To check the stability of the the differential Eq. \eqref{e8}, a Lyapunov function $ V(w) = \frac{1}{2} \sum_i { f_i(w)^2 } $ (a function of $ w $) is defined. As Lyapunov function directly or indirectly depends only on one control input i.e, window size hence, quadratic form of the function $ f_i(w) $ may provide the solution.  The following theorem depends on Eq. \eqref{e8} and states that $ w_i(t) $ converges to the unique stable point, where the optimal window size, $ w_i^* $ and optimal rate, $ x_i^* $  solves the aggregate system utility $  \sum_{i \in R} U_i (x_i) $.

\begin{thm}
Let $ V(w) = \frac{1}{2} \sum_i { f_i(w)^2 } $\\
$ V(w) $ is a Lyapunov function for the system of differential Eq. \eqref{e8}. The unique value minimizing $ V(w) $ is stable point of this system where all trajectories converge. \label{th1}
\end{thm}
\hspace*{-0.5cm}\textbf{Proof:}
\\
A Lyapunov function is a convex function and hence, the stable value of window size $ w(t) $ may lie either on boundary or within interior region. $ w(t) $ is an interior point if for any small value of $ \epsilon > 0 $, $ w(t+\epsilon) $ must lie in the same region as $ w(t) $ otherwise, $ w(t) $ is considered as boundary point. Now consider both the cases.
\\
\textbf{Case 1:} Interior point
\begin{equation}
\begin{aligned}
\frac{dV(w(t))}{dt} ={} & \frac{1}{2} \sum_j \frac{dV}{dw_j} \frac{dw_j(t)}{dt} \\
 	 	={} & \frac{1}{2} \sum_j \sum_i \frac{d (f_i(w))^2}{d w_j} \dot{w_j} \\
 	={} & -k \sum_j \sum_i f_i(w) \frac{df_i}{w_j} \dot{w_j}\\
 	={} & -k f(w)^T J_f \dot{w} .\label{e15}
 	\end{aligned}
\end{equation}
Where, $ J_f = \frac{df}{dw} = (d_{trans} + U') D^{-2} A (A^T X D^{-1} A)^{-1} A^T D^{-1}$ (Jacobian of function $ f $ with respect to $ w $ and proof is given in Appendix A). $ \dot{w} = f(d) f(w) = d_{trans} D^{-1} f(w) $. By using all these values, Eq. \eqref{e15} can be written as,
\begin{equation}
\begin{aligned}
\frac{dV(w(t))}{dt} ={} & -k f(w)^T \left( (d_{trans} + U') D^{-2} A (A^T X D^{-1} A)^{-1} A^T D^{-1} \right) \left(d_{trans} D^{-1} f(w) \right)\\
 	={} & -k f(w)^T \left[ (d_{trans} + U') D^{-2} A (A^T X D^{-1} A)^{-1} A^T D^{-1} \right]\left(d_{trans} D^{-1} f(w) \right)\\
 	={} & -k f(w)^T \left[ (d_{trans} + U') D^{-2} A A^{-1} D X^{-1} (A^T)^{-1} A^T D^{-1}\right]\left(d_{trans} D^{-1} f(w) \right)\\ 
 	={} & -k f(w)^T \left[ (d_{trans} + U') D^{-2} X^{-1} d_{trans} D^{-1} \right]  f(w) \\ 
 	={} & -k f(w)^T Q f(w). \label{e16}
 	\end{aligned}
\end{equation}
Here, $ Q = \left[ (d_{trans} + U') D^{-2} X^{-1} d_{trans} D^{-1} \right] $, $ d_{trans} = diag(d_{trans}^i(t)) $, $ U' = diag( U'_i(x_i(t))) $, $ D = diag(D_i(t)) $ and $ X= diag(x_i(t)) $. 

The inverse of a rectangular matrix cannot be calculated. Hence, first we need to convert matrix $ A $ into a square matrix by adding some vectors from a basis for $ \mathbb{R}^{RE} $. The matrix, $ Q $ in the bracket of Eq. \eqref{e16} is positive definite (the proof is provided in Appendix B). The overall value of Eq. \eqref{e16} is negative and hence, $ V(w(t)) $ is strictly decreasing in $ t $ on interior points.

\textbf{Case 2:} Boundary point\\
Jacobian matrix $ J_x $ can be defined only for the interior points but it may be extended to boundary points in the form of feasible direction $ \bar{d} $. A vector $ \bar{d} \in \mathbb{R}^n, \bar{d} \neq 0 $, is called a feasible direction at time $ t $ if there exist $ \epsilon_0>0 $ such that $ t $ and $ t+\epsilon d $ belong to same bottleneck set for all $ \epsilon \in [0, \epsilon_0] $.
\vspace*{-0.25cm}

\section*{Results and Analysis}
The simulation is set by constructing the time varying network according to the TVCN model proposed in Section 3.1. The proposed TVCN model is an example of real world networks, hence, degree distribution must follow power law distribution, $ P(k) \sim k^{-\alpha} $, where $ 2 < \alpha \leq 3 $. The parameters are set to be seed node $ n_0 = 5 $, number, $ M = 5 $, fraction of newly added links $ \beta $ in the range $ (0,1) $, fraction of rewired links $ \gamma $ is in the range $ (0.5, 1) $, with network size ranging from $ N = s \times 10^2 $ to $ N = 3.5 \times 10^3 $. Any node may be in the user's $ (S,D) $ sets or participate in routing also. Data forwarding capacity of a node $ n $, $ C_n $ is defined by the degree $ k_n $ of the node $ n $. Capacity of a link $ C_{e_{mn}} $ is obtained by multiplying the degrees $ k_m $ and $ k_n $ of end nodes $ m $ and $ n $, respectively. At each time stamp, degree of the nodes will be different, hence capacity of the nodes as well as links change accordingly. We demonstrate the convergence of users' parameters ($ w(t), (f(w))^2 $) with utility function, $ U_i(x_i(t)) = a_i log (x_i(t) + b_i) $, $ a_i > 0 $ and $ 0 \leq b_i \leq 1 $, through simulations and study the behavior of $ w^* $ for time varying network. We study the behavior of all the window update (La's ans Mo's) approaches on the networks designed by the time varying communication network model and compare with the proposed approach.

While the connection is established between user's source and destination pairs, initial value number of packets in window, $ w(0) $ is calculated by summing up the packets which are in transit as well as in queue on the resources which appear in user's route. Window size is adjusted with time and finally it converges to unique stable point $ w^* $. Simulation is run for $ 15000 $ iteration and convergence of window size for four users are mentioned in Table \ref{tab1}. The value of scaling parameter, $ k $, may be any positive number. Here, in the paper it is fixed as $ 0.1 $ and each user's data sending amount is fixed as $ 10Mb $. $ w(0) $ of User $ 1-4 $ are evaluated  as $ 25.1012\mbox{ } Mb, 15.2063\mbox{ } Mb, 20.4796\mbox{ } Mb$ and $ 35.6423 \mbox{ } Mb$, respectively. These values are taken as input for all approaches (the proposed, La's \cite{la} and Mo's \cite{mo2000fair}). In the proposed method, the value of $ w_i(t) $ depends on the delays ($ d_{prop}^i $, $ d_Q^i(t) $ and $ d_{trans}^i(t) $), $ s_i(t) $, willingness to pay $ p_i(t) $ and previous window size $ w_i(t-1) $ for user $ i $. $ w_i(t) $ will decrease exponentially with time and the value of $ w_i(t) $ makes $ s_i(t) \approx 0 $, is termed as optimal window size $ w_i^* $ of the user $ i $. As the window updates the equations for all the approaches are different hence, convergence rate will be different with difference in the value of $ w_i^* $. Mo's approach gives the value of $ w^* $ for all the users; User $ 1-4 $ as $ 3.2378\mbox{ } Mb, 15.2946\mbox{ } Mb, 5.2962\mbox{ } Mb$ and $ 9.9753\mbox{ } Mb$, respectively. La's approach gives the value of $ w^* $ of all users; User $ 1-4 $ as $ 1.3471\mbox{ } Mb, 7.7192\mbox{ } Mb, 2.0598\mbox{ } Mb$ and $ 2.0986\mbox{ } Mb$, respectively. By using the proposed approach, User $ 1-4 $ get $ w^* $ values as $ 1.3568\mbox{ } Mb, 7.8302\mbox{ } Mb, 2.0746\mbox{ } Mb$ and $ 2.1107\mbox{ } Mb$, respectively. Window update equation keeps changing dynamically with the varying value of transmission delay, $ d_{trans}(t) $ unlike, propagation delay, $ d_{prop} $ in La's and Mo's, which was static. The proposed approach provides a unique stable value of window size, $ w^* $ as in La's approach and is quite approximate to the value obtained through static (La's) approach. Mo's approach uses both propagation and queuing delays as static and it assumes less number of packets in queue irrespective of large number of incoming packets in the network hence, provides higher value of $ w^* $. The proposed method provides dynamicity in the previous approaches by updating the window size based on transmission delay and gets stable value of window size, $ w^* $.

\begin{table}
\caption{Comparative study of the convergence of values of window size, $ w $ obtained through proposed, Mo's and La's approaches, when network size $ N = 2500 $.} 
 \center
 \adjustbox{max height=\dimexpr\textheight-5.5cm\relax,
           max width=\textwidth}{
\begin{tabular}[1]{|l| c| c| c| c| c| c| c| c| c| c| c| c|}
\hline\hline
\textbf{Time} &  \multicolumn{4}{c|}{\textbf{Proposed}} & \multicolumn{4}{c|}{\textbf{Mo}} & \multicolumn{4}{c|}{\textbf{La}}\\[0.5ex]
\hline
& \textbf{User1} & \textbf{User2} & \textbf{User3} & \textbf{User4} &\textbf{User1} & \textbf{User2} & \textbf{User3} & \textbf{User4} & \textbf{User1} & \textbf{User2} & \textbf{User3} & \textbf{User4} \\
\hline\hline
$ 0 $ & $ 25.1012 $ & $ 15.2063 $ & $ 20.4796 $ & $ 35.6423 $ & $ 25.1012 $ & $ 15.2063 $ & $ 20.4796 $ & $ 35.6423 $ & $ 25.1012 $ & $ 15.2063 $ & $ 20.4796 $ & $ 35.6423 $  \\ 
\hline
$ 1000 $ & $ 11.3987 $ & $ 13.0505 $ & $ 10.7814 $ & $ 22.7718 $ & $ 7.8069 $ & $ 15.2946 $ & $ 8.8397 $ & $ 24.1478 $ & $ 4.6878 $ & $ 10.5919 $ & $ 4.0879 $ & $ 14.3736 $   \\
\hline
$ 2000 $ & $ 2.9560 $ & $ 11.3053 $ & $ 4.7454 $ & $ 11.8962 $ &  $ 3.2514 $ & $ 15.2946 $ & $ 5.4114 $ & $ 14.9804 $ & $ 1.3500 $ & $ 8.5821 $ & $ 2.0806 $ & $ 2.8991 $   \\
\hline
$ 3000 $ & $ 1.4584 $ & $ 10.1034 $ & $ 2.6813 $ & $ 5.6193 $ & $ 3.2379 $ & $ 15.2946 $ & $ 5.2987 $ & $ 11.0237 $ & $ 1.3471 $ & $ 7.9628 $ & $ 2.0600 $ & $ 2.1177 $  \\
\hline
$ 4000 $ & $ 1.3622 $ & $ 9.2974 $ & $ 2.1964 $ & $ 3.1047 $ & $ 3.2378 $ & $ 15.2946 $ & $ 5.2964 $ & $ 10.1366 $ & $ 1.3471 $ & $ 7.7866 $ & $ 2.0598 $ & $ 2.0992 $   \\
\hline
$ 5000 $ & $ 1.3571 $ & $ 8.7681 $ & $ 2.0983 $ & $ 2.3619 $ & $ 3.2378 $ &  $ 15.2946 $ & $ 5.2963 $ & $ 9.9978 $ & $ 1.3471 $ & $ 7.7377 $ & $ 2.0598 $ & $ 2.0986 $  \\
\hline
$ 6000 $ & $ 1.3568 $ & $ 8.4257 $ & $ 2.0791 $ & $ 2.1720 $ & $ 3.2378 $ &  $ 15.2946 $ & $ 5.2962 $ & $ 9.9784 $ & $ 1.3471 $ & $ 7.7243 $ & $ 2.0598 $ & $ 2.0986 $  \\
\hline
$ 7000 $ & $ 1.3568 $ & $ 8.2067 $ & $ 2.0754 $ & $ 2.1255 $ & $ 3.2378 $ & $ 15.2946 $ & $ 5.2962 $ & $ 9.9757 $ & $ 1.3471 $ & $ 7.7206 $ & $ 2.0598 $ & $ 2.0986 $  \\
\hline
$ 8000 $ & $ 1.3568 $ & $ 8.0674 $ & $ 2.0747 $ & $ 2.1143 $ & $ 3.2378 $ & $ 15.2946 $ & $ 5.2962 $ & $ 9.9753 $ & $ 1.3471 $ & $ 7.7196 $ & $ 2.0598 $ & $ 2.0986 $   \\
\hline
$ 9000 $ & $ 1.3568 $ & $ 7.9794 $ & $ 2.0746 $ & $ 2.1116 $ & $ 3.2378 $  & $ 15.2946 $ & $ 5.2962 $ & $ 9.9753 $ & $ 1.3471 $ & $ 7.7193 $ & $ 2.0598 $ & $ 2.0986 $ \\
\hline
$ 10000 $ & $ 1.3568 $ & $ 7.9239 $ & $ 2.0746 $ & $ 2.1109 $ & $ 3.2378 $ & $ 15.2946 $ & $ 5.2962 $ & $ 9.9753 $ & $ 1.3471 $ & $ 7.7192 $ & $ 2.0598 $ & $ 2.0986 $ \\
\hline
$ 11000 $ & $ 1.3568 $ & $ 7.8890 $ & $ 2.0746 $ & $ 2.1108 $ & $ 3.2378 $& $ 15.2946 $ & $ 5.2962 $ & $ 9.9753 $ & $ 1.3471 $ & $ 7.7192 $ & $ 2.0598 $ & $ 2.0986 $ \\
\hline
$ 12000 $ & $ 1.3568 $ & $ 7.8446 $ & $ 2.0746 $ & $ 2.1107 $ &$ 3.2378 $ & $ 15.2946 $ & $ 5.2962 $ & $ 9.9753 $ & $ 1.3471 $ & $ 7.7192 $ & $ 2.0598 $ & $ 2.0986 $ \\
\hline
$ 13000 $ & $ 1.3568 $ & $ 7.8302 $ &$ 2.0746 $ & $ 2.1107 $  &$ 3.2378 $ & $ 15.2946 $ & $ 5.2962 $ & $ 9.9753 $ & $ 1.3471 $ & $ 7.7192 $ & $ 2.0598 $ & $ 2.0986 $  \\
\hline
$ 14000 $ & $ 1.3568 $ & $ 7.8302 $ &$ 2.0746 $ & $ 2.1107 $  &$ 3.2378 $ & $ 15.2946 $ & $ 5.2962 $ & $ 9.9753 $ & $ 1.3471 $ & $ 7.7192 $ & $ 2.0598 $ & $ 2.0986 $  \\
\hline
$ 15000 $ & $ 1.3568 $ & $ 7.8302 $ &$ 2.0746 $ & $ 2.1107 $  &$ 3.2378 $ & $ 15.2946 $ & $ 5.2962 $ & $ 9.9753 $ & $ 1.3471 $ & $ 7.7192 $ & $ 2.0598 $ & $ 2.0986 $  \\
\hline
\end{tabular}
\label{tab1}
}
\end{table}

In the communication networks, multiple paths are available for sending packets for each user, between the desired source and destination. Among all the available paths, users select the shortest path to reach the destination with a maximum flow rate of individual links for data communication. The data sending rate $ x $ is reduced as multiple users want to share the common resources. With the varying size of the network $ N $ (from $ N = 5 \times 10^2 $ to $ N = 3.5 \times 10^3 $), shortest route of the user will also change. 
Routes of the users may be same or vary at different time instants. It depends on the availability of the resources. 

User's window size depends on two parameters; total number of packets in transit and queue. Using window update theorem (Eq. \eqref{e4}), $ w^* $ of each user is shown in Table \ref{tab3}. The size of the network, $ N $ is varying but the number of users is kept same as in initial network. Here, initial size of the network is considered as $ N = 5 \times10^2 $ and it ranges up to $ N = 3.5 \times 10^3 $ where, $ \Delta N = 5 \times 10^2 $ (Table \ref{tab3}).  The size of the network increases but $ w(t) $ decreases to a stable point $ w^* $. User's window size depends on the demand of particular resources appearing in the shortest route. If demand is high then window size and data sending rate will be less. When $ N $ is changing then, user's route, data flow rate $ x $ through the path and $ w^* $ will also change accordingly. Optimal window size, $ w^* $ for users are independently changing and in some cases $ w^* $ obtained through proposed approach is higher/lower than La's. But, Mo's approach uses more static approach and analyze less congestion in the network. 
\begin{table}
\caption{Comparative study of the values of optimal window size $ w^* $ obtained through proposed, La's and Mo's approaches, for different values of network size $ N $.} 
 \center
 \adjustbox{max height=\dimexpr\textheight-5.5cm\relax,
           max width=\textwidth}{
\begin{tabular}[1]{|l| c| c| c| c| c| c| c| c|}
\hline\hline
\textbf{Users} & & \multicolumn{7}{c}{\textbf{Network Size $ (N) $}}\\
& \textbf{Approaches} & \textbf{$ 500 $} & \textbf{$ 1000 $} & \textbf{$ 1500 $} & \textbf{$ 2000 $} & \textbf{$ 2500 $} & \textbf{$ 3000 $} & \textbf{$ 3500 $}
\\ [0.5ex]
\hline\hline
& Proposed & $ 5.0157 $ & $ 9.0963 $ & $ 4.1568 $ & $ 5.4256 $ & $ 1.3568 $ & $ 8.7011 $ & $ 6.4785 $\\[-1ex]
\raisebox{1.5ex}{User 1} & La's & $ 4.9549 $ & $ 4.9549 $ & $ 4.1032 $ & $ 5.2499 $ & $ 1.3472 $ & $ 8.5732 $ & $ 6.0885 $\\
& Mo's & $ 12.5436 $ & $ 14.0656 $ & $ 16.0440 $ & $ 9.4742 $ & $ 3.2378 $ & $ 27.9147 $ & $ 4.8265 $\\
\hline
& Proposed & $ 2.2173 $ & $ 0.6899 $ & $ 2.6955 $ & $ 1.9178 $ & $ 7.8302 $ & $ 3.0075 $ & $ 0.9175 $\\[-1ex]
\raisebox{1.5ex}{User 2} & La's & $ 2.1135 $ & $ 2.1135 $ & $ 2.6547 $ & $ 1.7302 $ & $ 7.7192 $ & $ 2.9698 $ & $ 0.9153 $\\
& Mo's & $ 8.5779 $ & $ 5.2688 $ & $ 5.4462 $ & $ 1.1634 $ & $ 15.2946 $ & $ 12.0407 $ & $ 4.3771 $\\
\hline
& Proposed & $ 5.9961 $ & $ 2.7512 $ & $ 4.9292 $ & $ 7.8832 $ & $ 2.0746 $ & $ 1.8750 $ & $ 2.1889 $\\[-1ex]
\raisebox{1.5ex}{User 3} & La's & $ 5.9337 $ & $ 5.9337 $ & $ 4.8319 $ & $ 7.7533 $ & $ 2.0598 $ & $ 1.8367 $ & $ 1.2650 $\\
& Mo's & $ 12.0888 $ & $ 1.9225 $ & $ 11.3823 $ & $ 11.4005 $ & $ 5.2962 $ & $ 3.3067 $ & $ 8.2767 $\\
\hline
& Proposed & $ 0.8467 $ & $ 1.7196 $ & $ 5.0519 $ & $ 1.0817 $ & $ 2.1107 $ & $ 1.3798 $ & $ 1.2145 $\\[-1ex]
\raisebox{1.5ex}{User 4} & La's & $ 0.8455 $ & $ 0.8455 $ & $ 5.0073 $ & $ 1.0773 $ & $ 2.0986 $ & $ 1.3743 $ & $ 1.2112 $\\
& Mo's & $ 7.0850 $ & $ 13.6237 $ & $ 12.8835 $ & $ 6.8177 $ & $ 9.9753 $ & $ 6.3977 $ & $ 8.1358 $\\
\hline
\end{tabular}
\label{tab3}
}
\end{table}

\begin{figure}[!htb]
\begin{center}
\includegraphics[width=0.7\linewidth, height=2.25 in]{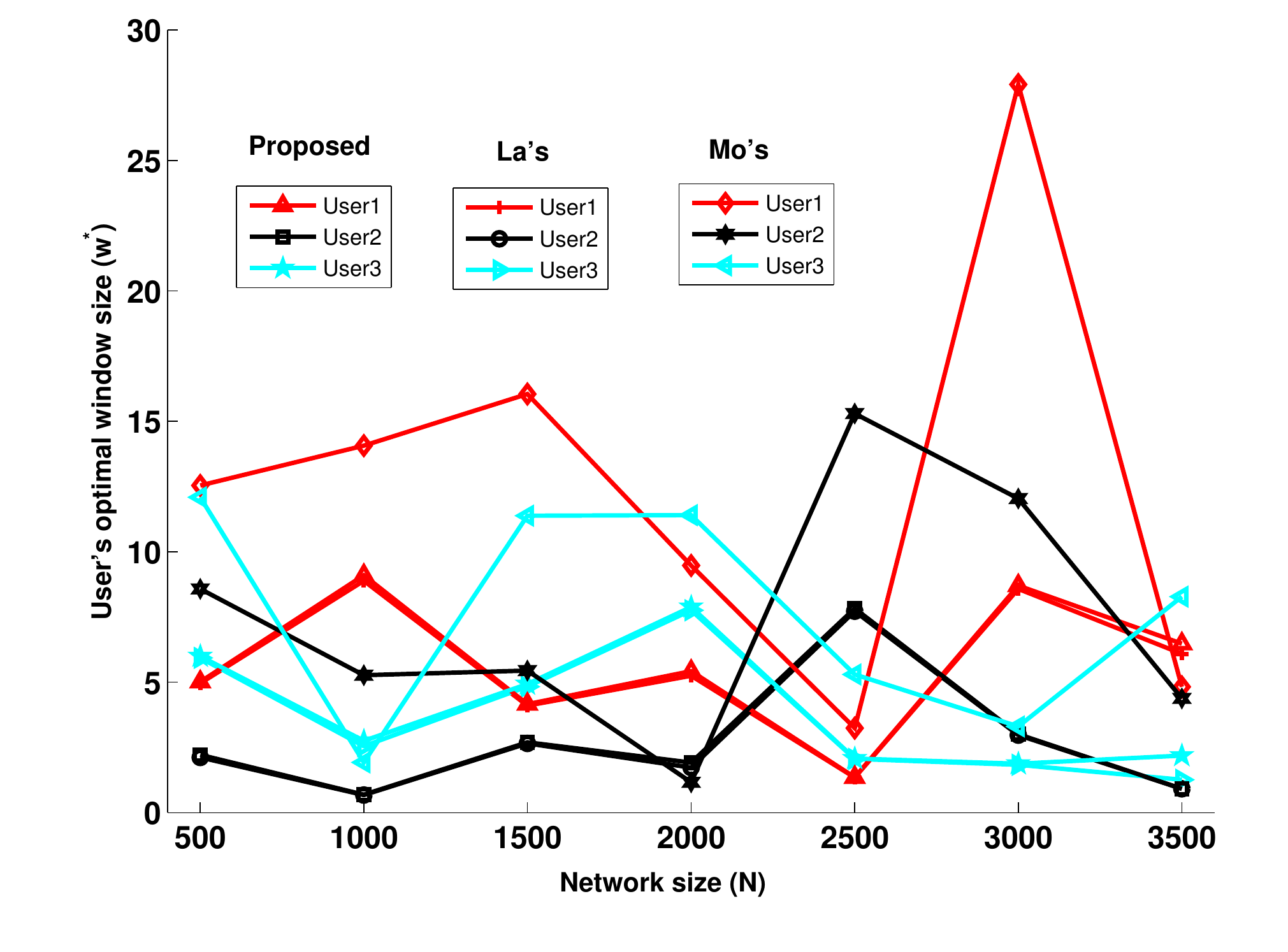} 
\caption{ Comparative study of the values of $ w^* $ obtained through proposed, La's and Mo's approaches, for the network size $ N = \{5 \times 10^2 $, $ 1 \times 10^3 $ ... , $ 3.5 \times 10^3\} $.}
\label{f2}
\end{center} 
\end{figure}

A comparative study of optimal window size, $ w^* $ of three users; User1, User2 and User3 are shown by using red, black and cyan lines with different markers (Figure \ref{f2}). La's and proposed approaches give approximate values of $ w^* $ while Mo's approach considers more static approach and gives higher value of $ w^* $ for each user.  

\begin{figure}[!htb]
\begin{center}
\includegraphics[width=0.7\linewidth, height=2.25 in]{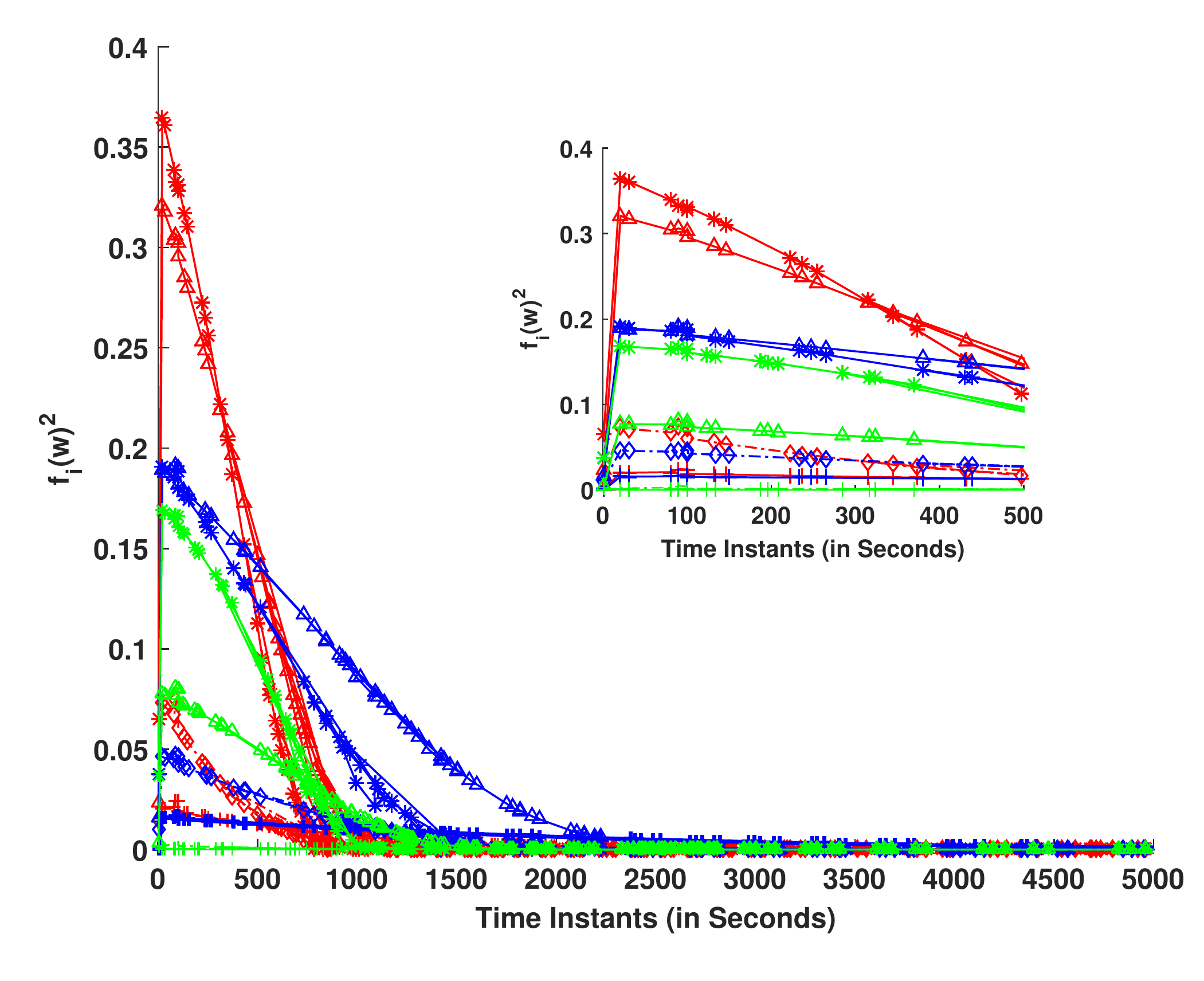} 
\caption{convergence of user's $ \frac{f_i(w)^2}{2} $ value, when $ N=3 \times 10^4 $.}
\label{f3}
\end{center} 
\end{figure}

The window size vector $ w(t) $ is said to be $ (p,1) $ proportionally fair rate vector $ x $ if $ s_i(t) \approx 0 $ for all $ i $ and the unique value minimizing $ V(w) $ is stable point of this system. Figure \ref{f3} shows convergence of $ V(w) = \frac{f_i(w)^2}{2} $ for $ 4 $ users when size of the network $ N = 3 \times 10^4 $. Initially, the difference between actual and targeted backlog, $ s_i $ for user $ i $ is high hence, $ V(w) $ is also high. But it decreases independently and exponentially with time and finally, it becomes approximately zero, like $ s_i $.

In window update equation (Eq. \eqref{e7}), La added dynamic terms ($ U'_i(x_i(t))+ x_i(t)U''_i(x_i(t)) $) to the propagation delay $ (d_{prop}) $ which will always be positive and depends on the dynamic rate and corresponding user's utility function. We have divided La's approach into two parts: (i) with the consideration of dynamic terms and (ii) without the consideration of dynamic terms (LaWD approach). It is found that both the approaches converge to the same value of stable window size, $ w^* $. But, the computation complexity in the convergence of window size is different for both the approaches and is shown in Table \ref{tab4}. The configuration of the system is given as Intel $ R $ Xeon (R) CPU E5-$ 2620 $ v2 @ $ 2.10 $ GHz $ 2.10 $ GHz with memory size $ 8.00 $ GB. The average value of the computation complexity for different values of the network size, $ N $ is calculated to check the speed up of the proposed approach with other approaches. Total computation time for obtaining  $ w^* $ depends upon the network topology and the total number of congested resources. Different network structures with the same number of nodes and resources (links) take a different amount of time to get optimal window size. The simulation is run for $ 10 $ times and finally, an average value is considered as computation time. The proposed approach is approximately $ 23, 33 $ and $ 24 $ times faster than the Mo's, La's and La without Dynamic terms (LaWD) approaches, respectively.

\begin{table}
\caption{Comparative study of the computation complexity (in Seconds) analysis to obtain $ w^* $ through proposed, Mo's, La's and LaWD's approaches, for different size of network $ N $.} 
 \center
 \adjustbox{max height=\dimexpr\textheight-5.5cm\relax,
           max width=\textwidth}{
\begin{tabular}[1]{|l| c| c| c| c|}
\hline\hline
\textbf{Network size $(N)$} & \textbf{Proposed} & \textbf{Mo} & \textbf{La} & \textbf{LaWD}\\[0.5ex]
& \multicolumn{4}{c}{\textbf{time taken (in seconds) for optimal window size $ (w^*) $}}\\
\hline\hline
$ 500 $ & $ 281.4865 $ & $ 1.2660 \times 10^3 $ & $ 2.9717 \times 10^3 $ & $ 2.5891 \times 10^3 $ \\ 
\hline
$ 1000 $ & $ 279.2723 $ & $ 7.4122 \times 10^3 $ & $ 2.9577 \times 10^3 $ & $ 2.7117 \times 10^3 $ \\
\hline
$ 1500 $ & $ 484.9320 $ & $ 6.7600 \times 10^3 $ & $ 1.3749 \times 10^4 $ & $ 1.3954 \times 10^4 $ \\
\hline
$ 2000 $ & $ 744.9612 $ & $ 2.2145 \times 10^4 $ & $ 3.6035 \times 10^4 $ & $ 1.2091 \times 10^4 $ \\
\hline
$ 2500 $ & $ 461.6515 $ & $ 1.2812 \times 10^4 $ & $ 1.8069 \times 10^4 $ & $ 1.5984 \times 10^4 $ \\
\hline
$ 3000 $ & $ 243.9001 $ & $ 1.3780 \times 10^4 $ & $ 9.3088 \times 10^3 $ & $ 8.4913 \times 10^3 $ \\
\hline
$ 3500 $ & $ 550.8467 $ & $ 6.9841 \times 10^3 $ & $ 1.6396 \times 10^4 $ & $ 1.6720 \times 10^4 $ \\
\hline
\end{tabular}
\label{tab4}
}
\end{table}

\vspace*{-0.25cm}
\section*{Conclusion and Future Directions}
A framework is designed to represent time-varying communication networks (TVCN). We have shown the existence of window based fair end to end congestion control scheme using a multi-class fluid based model. $ (p,1) $ proportionally fairness is used to formulate an optimization problem. According to the willingness for pay, $ (p_i) $ of each user, an optimal window size is allocated such that congestion is minimized as well as system utility is maximized. The window-based update algorithm is proposed by considering dynamic delay especially, transmission delay $ d_{trans}(t) $. It is observed that we are still getting a stable value of window size $ w^* $ which is approximately same as the result obtained through La's approach \cite{la}. While, Mo's approach \cite{mo2000fair} considers static delay hence, overestimates the value of optimal window size, $ W^* $. Convergence of the proposed window-based congestion control scheme is proved by using a Lyapunov function, $ V(w) $. For the same network topology, the converged window size, $ w^* $ is obtained in the least amount of time through the proposed scheme. The network is time-varying therefore, User's optimal window size, $ w^* $ is evaluated for the varying sizes of networks and we may get a different value of $ w^* $ due to change in route of each user.  

In future work, the proposed study will be extended to provide good QoS (Quality of Service) by considering error rates, robustness, data transmission within budget etc. The concept of multi-path routing can be used for data transmission of each user.  
\vspace*{-0.25cm}

\bibliographystyle{elsarticle-num}
\bibliography{letter} 

\begin{thebibliography}{10}
\bibitem{max}
D.~Nace, M.~Pi{\'o}ro, Max-min fairness and its applications to routing and
  load-balancing in communication networks: a tutorial, IEEE Communications
  Surveys \& Tutorials 10~(4).
\end{thebibliography}


\section*{Appendix A}
A Lyapunov function $ V(w) $ can be used to find an output error and/or size of a state, deviation from true parameter, energy difference from desired equilibrium point or it may be the combination of the above. Lyapunov function is decreasing with time (continuous or discrete) hence, we need to choose some suitable control equation such that $ \frac{d V(w(t))}{dt} \leq 0 $ and it should be minimum. 

\begin{thm}
Let $ V(w) = \frac{1}{2} \sum_i f_i(w)^2 $\\
$ V(w) $ is a Lyapunov function for the system of differential Eq. \eqref{e8}. The unique value minimizing $ V(w) $ is stable point of this system where all trajectories converge. \label{th1}
\end{thm}

Define
\begin{eqnarray}
&& J_x = \left[ \frac{\partial x_i}{ \partial w_j}, \mbox{    } i,j \in \mathcal{N} \right] , \nonumber \\
&& J_q = \left[ \frac{\partial q_i}{ \partial w_j}, \mbox{    } i \in \mathcal{L}, j \in \mathcal{N} \right]  \nonumber
\end{eqnarray}
$ J_x $ and $ J_q $ are considered as changes in rate and queue size with respect to the window size $ w $. All congested links are considered to solve the Lyapunov function $ V(w) $ and these congested links are collectively stored in a set $ B $. A sub-matrix $ A_B $ of matrix $ A $ is formed by considering only congested links.
\begin{lem}
The Jacobian $ J_x $ of $ x(w) $ with respect to $ w $ is given by following expression on the interior point: 
\begin{equation}
J_x = D^{-1}\left(I-XA_B(A_B^TXD^{-1}A_B)^{-1} A_B^T D^{-1}\right) \label{e9}
\end{equation}
\end{lem} 
\textbf{Proof:}
\begin{eqnarray}
&& D = diag \left[ (d_{P})_i+ A d_{trans}^i(t) + A(d_{Q}^t)_i \right] \nonumber\\
&& X = diag (x_i, \mbox{   } i \in \mathcal{N})\nonumber
\end{eqnarray}
To calculate $ J_x $, only congested links are considered and hence, the equation will be
\begin{eqnarray}
&& x_i \left[ (A_B {d_{Q}}_B)_i + (A_B {d_{T}}_B)_i + (d_{P})_i \right] = w_i , \mbox{   } i \in B  \label{e10}\\
&& A^T x = C \label{e11} 
\end{eqnarray} 
Now, taking partial derivative of Eq. \eqref{e10} with respect to $ w_j $ and avoiding subscript $ B $ for calculation
\begin{eqnarray}
&& \frac{\partial (x_i \left[ (A d_Q)_i + (A d_{trans})_i + (d_{prop})_i \right]}{\partial w_j} = \frac{\partial w_i}{w_j}\nonumber \\
&& \frac{\partial x_i}{\partial w_j} \left[ (A d_Q)_i + (A d_{trans})_i + (d_{prop})_i \right] + x_i \left[ A_i \frac{\partial (d_Q)_i}{\partial w_j} \right] = \delta_{ij}\nonumber \\
&& (D)_i (J_x)_{ij} + x_i A_i (J_q)_{ij} = \delta_{ij} \nonumber
\end{eqnarray}
Now this can be written in matrix form as
\begin{equation}
D J_x + XAJ_q  = I \label{e12}
\end{equation}
Multiplying both side by $ A^T D^{-1} $
\begin{equation}
A^T J_x + \left( A^T D^{-1}XA \right) J_q = A^T D^{-1} \label{e13}
\end{equation}
From Eq. \eqref{e11} $ A^T J_x = 0 $, hence
\begin{equation}
J_q = \left( A^T D^{-1}XA \right)^{-1} A^T D^{-1} \label{e14}
\end{equation}
Now, putting Eq. \eqref{e14} into Eq. \eqref{e12}, we get
\begin{eqnarray}
&& D J_x + XA \left( A^T D^{-1}XA \right)^{-1} A^T D^{-1}  = I \nonumber \\
&& J_x = D^{-1} \left( I - XA \left( A^T D^{-1}XA \right)^{-1} A^T D^{-1} \right) \nonumber
\end{eqnarray}

\begin{lem}
The Jacobian $ J_f $ of $ f(w) $ with respect to $ w $ is given by following expression on the interior point: 
\begin{equation}
J_f = (d_{trans} + U') D^{-2} A (A^T X D^{-1} A)^{-1} A^T D^{-1} \label{e10}
\end{equation}
\end{lem} 

\textbf{Proof:}
\begin{equation}
\begin{aligned}
J_f = {}& \frac{\partial f}{\partial w}=\frac{\partial \left(1 - \frac{x_i d_{trans}^i}{w_i} - \frac{x_i U'_i(x_i))}{w_i}\right)}{\partial w_j}\\
={}& -(J_x)_{ij} d_{trans}^i w_i^{-1} + x_i d_{trans}^i w_i^{-2} - (J_x)_{ij} U'_i(x_i) w_i^{-1} + x_i U'_i(x_i) w_i^{-2}. 
\end{aligned}
\end{equation}
Now, the above equation can be written in matrix form as,
\begin{equation}
\begin{aligned}
J_f ={}& X d_{trans} W^{-2} + X U'W^{-2} - \left( d_{trans} W^{-1} +  W^{-1} U' \right) J_x\\
={}& X d_{trans} W^{-2} + X U'W^{-2} - \left( d_{trans} W^{-1} +  W^{-1} U' \right) \left( D^{-1} - D^{-1} XA \left( A^T D^{-1}XA \right)^{-1} A^T D^{-1} \right)\\
={}& X d_{trans} W^{-2} + X U'W^{-2} - d_{trans} W^{-1} D^{-1} - W^{-1}U'D^{-1} + \left( d_{trans} + U'\right) W^{-1} D^{-1} XA \left( A^T D^{-1}XA \right)^{-1} A^T D^{-1} 
\end{aligned} \label{20}
\end{equation}
Where, $ X = WD^{-1} $. Hence, the starting four terms of Eq. \eqref{20} will be canceled out and it can be rewritten as, 
\begin{equation}
\begin{aligned}
J_f ={}& \left( d_{trans} + U'\right) W^{-1} D^{-1} XA \left( A^T D^{-1}XA \right)^{-1} A^T D^{-1}\\
={}& \left( d_{trans} + U'\right) D^{-2} A \left( A^T D^{-1}XA \right)^{-1} A^T D^{-1} \label{21}
\end{aligned} 
\end{equation} 
\section*{Appendix B}
\textbf{Proof for checking whether the matrix, $ Q $ is positive definite}
\begin{figure}[!htb]
\begin{center}
\includegraphics[width=0.5\linewidth, height=1.5 in]{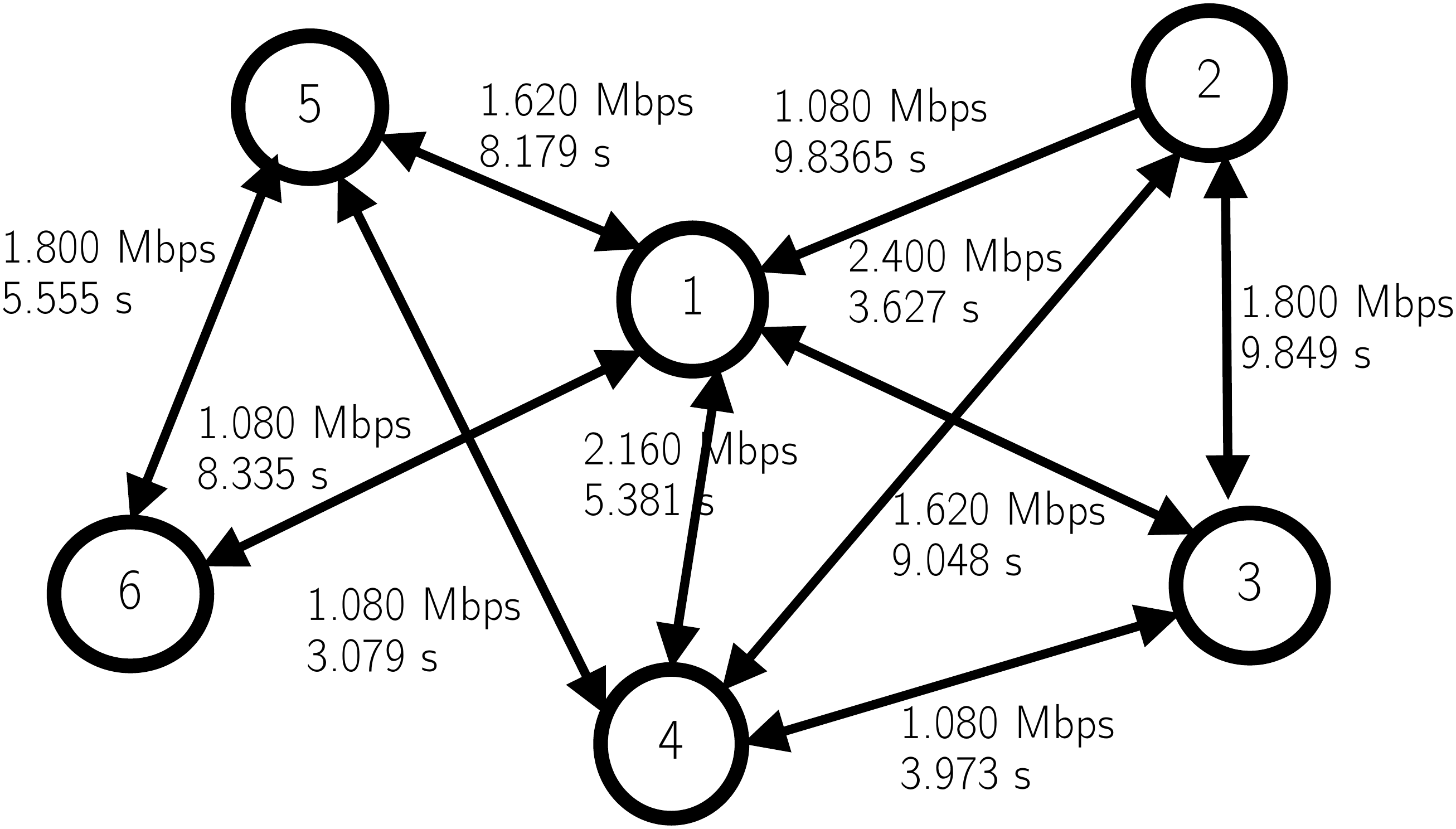} 
\caption{Example network}
\label{f4}
\end{center} 
\end{figure}

A numerical example of a simple network is given to check the matrix, $ Q $ (in the bracket of Eq. \eqref{e16}) is positive definite. There are three users who share the network. The capacities and the propagation delays of the resources are indicated next to the links. The source, destination, transmission delay, total delay, data rate, $ U' $ and user's route are provided in the Table \ref{tab5}.
\begin{table}
\caption{User's data.} 
 \center
\begin{tabular}[1]{|l| c| c| c| c| c| c| c|}
\hline\hline
User & Source & Destination &  Route & $ d_{trans}^i $ & $ U'_i(x_i) $ & $ x_i $ & $ D $ \\
 \hline
 User1 & $ 3 $ & $ 6 $ & $ 3-1-6 $ & $ 9.2593 $ & $ 0.6329 $ & $ 1.0800 $ & $ 42.0744 $ \\
 User2 & $ 1 $ & $ 4 $ & $ 1-4 $ & $ 4.6296 $ & $ 0.3759 $ & $ 2.1600 $ & $ 14.6398 $\\
 User3 & $ 6 $ & $ 2 $ & $ 6-5-4-2 $ & $ 9.2593 $ & $ 0.6329 $ & $ 1.0800 $ & $ 43.1729 $\\
 \hline 
\end{tabular}
\label{tab5}
\end{table}

By putting all these values, we get the matrix, $ Q $ as,
\[
\begin{bmatrix}
    0.0011       & 0 & 0  \\
    0       & 0.0034 & 0 \\
    0       & 0 & 0.0011.
\end{bmatrix}
\]
The eigen values of the above matrix are positive with values $ (0.0011, 0.0034 $ and $ 0.0011) $ hence, the matrix, $ Q $ is positive definite.
\begin{table*}
\caption{List of Symbols} 
 \center
 \adjustbox{max height=\dimexpr\textheight-5.5cm\relax,
           max width=\textwidth}{
\begin{tabular}[1]{l| p{110mm}} 
\hline\hline 
 Symbols & Meaning
\\ [0.5ex]
\hline\hline 
$ \Pi $ & Probability that a node  $ i $ will be selected through preferential attachment\\
\hline
$ \Pi' $ & Probability that a node  $ i $ will be selected through anti-preferential attachment\\
\hline
$ N $ & Set of nodes \\
\hline
$ E $ & Set of links  \\
\hline
$ T $ & Life span of the networks\\
\hline
$ x^\star $ & Optimal data rate \\
\hline
$ C_{e_{mn}} $ & Capacity of a link $ e_{mn} $ and $ e_{mn} \in E $ \\
\hline
$ A_{i,j,t_i} = 1 $ & If nodes $ i $ and $ j $ are connected at time $ t_i $ then the value of the matrix $ A $ will be $ 1 $ otherwise $ 0 $    \\
\hline
$ x_r(t_i) $ & Data flow rate of user $ r $ at time $ t_i $\\
\hline
$ U_{r,t_i}(x_r (t_i)) $ & System utility of user $ r $ with rate $ x_r (t_i) $.\\
\hline
$\psi_{e_{mn}}(t_i) $ &  total data flow through a link $ e_{mn} $ at time $ t_i $ \\
\hline
$\mathcal{P}_r(t_i)$ & willingness to pay of User $ r $ at time $ t_i $ \\
\hline
$ SYSTEM(U(t_i),A(t_i)) $ & Aggregate System Utility \\
\hline
$ d_{prop} $ & propagation delay \\
\hline
$ d_Q $ & queuing delay \\
\hline
$ d_{trans} $ & transmission delay\\
\hline
$ D $ & total delay\\
\hline 
$ w_i(t) $ & window size of user $ i $ at time $ t $\\
\hline
$ x_i(t) $ & data flow rate of user $ i $ at time $ t $\\
\hline
$ s_i $ & actual backlog of packets at user $ i $\\
\hline
$ n_0 $ & Initial number of nodes in the seed networks.\\
\hline
$ M (\leq n_0) $ & A number is selected for network expansion, rewiring and removal of links.\\
\hline
$ \sigma(S \rightarrow D) $ & Shortest path between $ S $ to $ D $. \\
\hline
$ V(w) $ & Lyapunov function \\
\hline
\end{tabular}
\label{tab6}
}
\end{table*}

\end{document}